\documentclass[a4paper]{article}

\usepackage{INTERSPEECH2021}
\usepackage{graphicx}
\usepackage{mathrsfs}
\usepackage{multirow}
\usepackage{tabularx}
\usepackage{lipsum}

\newcommand\blfootnote[1]{%
	\begingroup
	\renewcommand\thefootnote{}\footnote{#1}%
	\addtocounter{footnote}{-1}%
	\endgroup
}

\newcommand{\computecost}{q}
\newcommand{\bandwidth}{\mu}
\newcommand{\framerate}{\rho}
\newcommand{\matdimx}{w}
\newcommand{\matdimy}{v}
\newcommand{\steps}{m}
\newcommand{\framestep}{t}

\title{Amortized Neural Networks for Low-Latency Speech Recognition}
\name{Jonathan Macoskey, Grant P. Strimel, Jinru Su, Ariya Rastrow }

\address{
  Alexa Machine Learning, Amazon, USA}
\email{\{macoskey, gsstrime, jinru, arastrow\}@amazon.com}

\begin{document}

\maketitle

\begin{abstract}
We introduce Amortized Neural Networks (AmNets), a compute cost- and latency-aware network architecture particularly well-suited for sequence modeling tasks. 
We apply AmNets to the Recurrent Neural Network Transducer (RNN-T) to reduce compute cost and latency for an automatic speech recognition (ASR) task. 
The AmNets RNN-T architecture enables the network to dynamically switch between encoder branches on a frame-by-frame basis. 
Branches are constructed with variable levels of compute cost and model capacity. 
Here, we achieve variable compute for two well-known candidate techniques: one using sparse pruning and the other using matrix factorization. 
Frame-by-frame switching is determined by an arbitrator network that requires negligible compute overhead. 
We present results using both architectures on LibriSpeech data and show that our proposed architecture can reduce inference cost by up to 45\% and latency to nearly real-time without incurring a loss in accuracy. 
\blfootnote{First two authors contributed equally.}

\end{abstract}
\noindent\textbf{Index Terms}: automatic speech recognition, latency reduction, compute cost optimization, on-device

\section{Introduction}

Production ASR systems using end-to-end, neural architectures such as the RNN-T \cite{Graves2013, Rao2018, Kannan2019}, have enabled the deployment of generalized ASR systems in compute constrained settings. 
While ASR systems are under an ever-increasing demand to perform in a more natural, responsive manner in a multitude of domains, there has also been an increased desire to enable this technology to execute locally on lower power edge devices where compute resources are limited. 
When using lower-performance hardware that is unable to perform ASR in real time, a backlog of audio data builds and can cause significant user-perceived latency (UPL) \cite{Sainath2020}.

A simple approach for minimizing UPL caused by resource constraints is to reduce ASR model size such that the compute cost of a model does not throttle local processing performance.
Many strategies have been proposed for neural ASR model size reduction such as moving from the LSTM \cite{Hochreiter1997} to more efficient layers including the CIFG-LSTM \cite{Greff2017}, LSTM-p \cite{Jia2017}, or Simple Recurrent Unit (SRU) \cite{Lei2018}. 
Each of these architectural changes use fewer parameters while maintaining accuracy. Other non-architectural methods of model size reduction include sparse pruning \cite{Zhu2018,Shangguan2019}, matrix factorization \cite{Pang2018}, weight quantization \cite{Alvarez2016,Nguyen2020}, and knowledge distillation \cite{Panchapagesan2020}. 
These strategies becomes less attractive when the computational constraints of ever-shrinking edge devices mandate extreme levels of model compression while society's demands for ASR are constantly expanding and require increasing model capacity.

During real-time ASR decoding, utterances are processed as a sequence of feature frames that can vary wildly in their inherent complexity. 
For example, silent frames often exist during pauses between acoustically rich words and phonemes as well as before and after the main content of an utterance. 
Despite this variability, typical speech processing systems use a predetermined network with a fixed compute cost and model capacity to evaluate each individual frame;
this is a missed opportunity to accelerate ASR by dynamically varying the amount of compute power used throughout an utterance. 
Prior studies have established the applicability of variable approaches in other domains including natural language processing \cite{Tyagi2020}, character-level prediction and language modeling \cite{Bengio2015b,Graves2016}, music modeling \cite{Jernite2016}, and vision \cite{Bolukbasi2017}.

In this work, we present a new neural network architecture for speech processing that amortizes the cost of processing complex frames over an entire sequence of frames to reduce latency. 
We apply our AmNets approach to the RNN-T architecture for ASR applications. 
Here, we bifurcate the RNN-T encoder into two branched encoder networks in which the branches are compressed to different degrees to achieve varying levels of compute cost. 
For each frame, a recurrent arbitrator network is trained to decide which encoder branch to use. 
We optimize this decision making process by introducing a novel amortized loss function that trains the arbitrator to deliberately choose variable compute levels depending on the instantaneous accuracy-latency trade off within a given sequence.
Prior methods utilizing adaptive compute have instead focused on reducing \emph{average} cost, which acts as an indirect regularizer for real-time speech processing latency.
We show that this is sub-optimal in comparison to our architecture, which is driven by a direct optimization of the real-time audio processing backlog. 
We illustrate the efficacy and flexibility of the proposed methods by reporting results from two separate branch architectures using sparse pruning and matrix factorization.

The key contributions of this paper are the amortized recurrent neural network architecture (Section \ref{sec:amnet_architecture}), the application of AmNets to on-device, low latency speech processing (Section \ref{sec:candidates}), and the amortized latency loss function (Section \ref{sec:am_latency_loss}). Section \ref{sec:experiments} discusses experimental details and results.

\section{Amortized Network Architecture}
\label{sec:amnet_architecture}
\begin{figure*}[th!]
	\centering
	%\vspace{-2.0mm}
	\includegraphics[width=\linewidth]{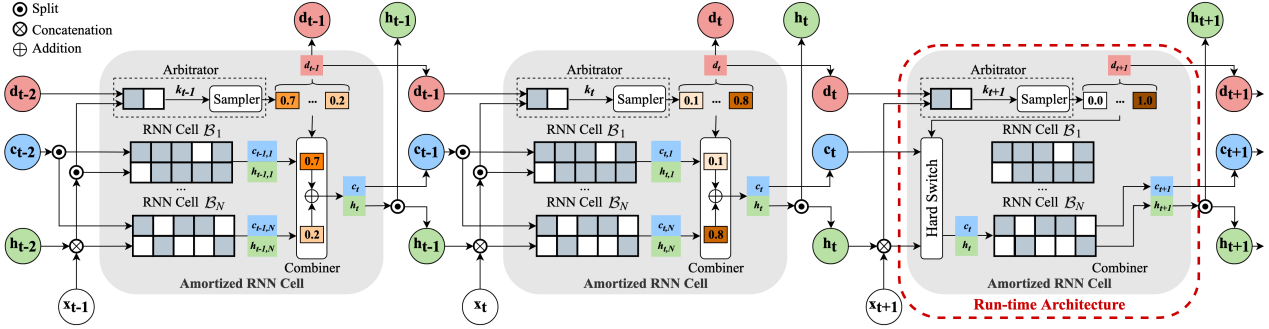}
	%\vspace{-2.0mm}
	\caption{\small Unfolded AmRNN layer with $N$ branches. Each block represents the execution of an AmRNN layer at a single time step, i.e., a single cell. Left and center blocks illustrate branch combining during training; right block shows run-time hard branch switching.}
	%\vspace{-4.0mm}
	\label{fig:architecture}
\end{figure*}

The fundamental component of an AmNet is the Amortized RNN (AmRNN) layer shown in Figure \ref{fig:architecture}.
An AmRNN layer consists of an arbitrator network, RNN cell branches $\mathcal{B}_1, \ldots, \mathcal{B}_N$, and a branch state combiner.
The arbitrator is the central mechanism for gating between branches, and it is implemented as an RNN so that it is able to maintain knowledge of prior decisions.
The arbitrator is much smaller in size than any of the RNN cell branches such that it does not add noticeable compute burden at inference time.
The arbitrator takes an input vector of pre-processed audio feature frames $x_{t}$ and computes output branch arbitration decisions $k_{t}=[\pi_{1}, \ldots, \pi_{N}]$.
The arbitrator can optionally take prior branch output states $h_{t-1}$ and prior arbitration decisions $d_{t-1}$.
$k_{t}$ is passed to a sampler, which is responsible for converting those decisions to a harder decision $d_{t} \in [0, 1]$ sampled from $k_{t}$.
During training, we use the Gumbel-Softmax reparameterization trick to compute $d_{t}$
\noindent
\begin{equation}\label{eqn:gumbel}
	\noindent
	d_{t}(n)=\frac{\textrm{exp}((\mathrm{log}(\pi_n)+g_n)/\tau)}{\sum_{j=1}^{N}\textrm{exp}((\mathrm{log}(\pi_j)+g_j)/\tau)},
\end{equation}
\noindent
which allows gradients to flow through non-differentiable stochastic nodes during backpropagation \cite{Jang2017}; in this case, a discrete categorical distribution. 
In Equation \ref{eqn:gumbel}, $g_1,...,g_N$ are i.i.d. samples taken from the Gumbel distribution \cite{Gumbel1960}.
Hyperparameter $\tau \in (0,\infty)$ controls the discreteness of the one-hot sampling, where values of $\tau$ close to zero approach perfectly hard sampling.
Throughout training, $\tau$ can be gradually decreased to coerce the network towards more discrete decisions.
During run-time, the sampler is switched to one-hot encoding such that only one branch is computed for each frame. 

AmRNN branches can be comprised of any recurrent cells such as LSTM, GRU, and SRU and can be of mixed dimensionality to achieve a balance between model capacity and compute efficiency across branches. 
Branches can also consist of stacked RNNs to increase model depth. 
Here, we only discuss AmNets with equal state sizes across all branches. 
However, mixed dimensionality can be achieved by including dense layers between branches to linearly project branch states into the dimensions of other branches \cite{Macoskey21}.

During training and at every time step $\framestep$, each branch receives prior AmRNN states $\{h_{t-1}, c_{t-1}\}$ and input features $x_{t}$ and produces output states $\{h_{t,n}, c_{t,n}\}$ for each branch. These states are passed to the combiner, which uses sampler decisions $d_{t}$ to return a linear combination of states from each branch as a single set of output states $\{h_t, c_t\}$ for each time step:
\noindent
\begin{align}\label{eqn:states}
	\noindent
	\{h_t, c_t\} = \sum_{n=1}^{N} d_{t-1}(n) \{h_{t,n},c_{t,n}\}
\end{align}
\noindent 
During run-time, the decision sequence from the sampler is treated as a hard switch such that only one branch receives an input and produces an output for the entire cell.

The AmRNN outputs branch decisions and associated costs to compute an amortized loss $\mathcal{L}_{\text{compute}}$ over an entire sequence.

\section{Candidate RNN-T Architectures}
\label{sec:candidates}
The RNN-T is an end-to-end neural architecture that is well-suited for sequence-to-sequence modeling applications and has shown recent success for ASR modeling, especially on-device \cite{He2019}.
This architecture consists of three main components: two multi-layer RNNs, the encoder $\mathcal{F}$ and decoder $\mathcal{G}$, and a joint network $\mathcal{J}$.
For ASR, $\mathcal{F}$ takes a vector of pre-processed audio feature frames $x_{1:T}$ as input and returns output $h_t^{enc}$. 
$\mathcal{G}$ takes a vector embedding of previous non-blank labels $y_{1:M}$ and returns intermediate output $h_m^{dec}$.
Outputs from the encoder and decoder are passed to the joint network.
Here, $\mathcal{J}$ is implemented as an additive operation between encoder and decoder outputs but can also include a joint feedforward network.
The conditional output distribution $P\left(\hat{y}_{m+1}|x_{1:t}, y_{1:m}\right)$ is obtained by applying a softmax to the result of $\mathcal{J}$. 

Typically, the encoder network is significantly larger than the decoder.
Additionally, during run-time, the decoder is cacheable over states with the same partial sub-word sequence and lends itself to batched execution (with multi-threading).
We therefore use AmNets to reduce encoder cost since it is the primary latency bottleneck at run-time. 
In all AmNets in this study, we opt for two branches: one with a higher cost and capacity, and one with a lower cost and capacity. We demonstrate the flexibility of AmNets by showcasing two candidate implementations, each using a different compression scheme to achieve varied compute cost between branches.

In the first candidate architecture, we compress branch weights using sparse pruning. Sparsity-inducing regularization of neural networks has shown to achieve lowered compute costs while maintaining accuracy \cite{Wu2020-Sparsity, Hastie2015, Zhou2019, Gale2019, Hebiri2020}. In this work, we achieve various encoder sparsity levels by gradually sparse pruning weights throughout training \cite{Zhu2018}. Given an initial sparsity ratio of 0, we achieve a final sparsity ratio of $s_f$ over $\nu$ pruning steps starting at step 0 with a pruning frequency of $\Delta \steps$:
\noindent
\begin{align}\label{eqn:sparse_pruning}
	\noindent
	s_t = s_f - s_f\left (1 - \frac{\steps}{\nu\Delta \steps}\right )^3, m \in \{0, \Delta \steps, \ldots, \nu \Delta \steps\}
\end{align}
\noindent
Sparsity is enforced by including binary sparse masks alongside network weight matrices within each layer. 
For example, a single LSTM cell undergoing sparse pruning would include corresponding masks for the cell, input gate, output gate, and forget gate matrices of equivalent dimension. 
Throughout training, sparsity levels defined by Equation \ref{eqn:sparse_pruning} are applied to sparse masks in a point-wise fashion based on the magnitudes of model weights. 
At time $t$, the mask indices set to zero correspond to the indices of weights with magnitude closest to zero based on the ratio defined by $s_t$. 
At the beginning of each training batch, weight matrices are point-wise multiplied by their masks to set pruned weights to zero on each forward pass. 
By using sparse masks, masked model weights are not dropped; hence, this policy enables the model to recover from early bad masking decisions.
For AmNets, separate sparsity trackers are used for each branch to achieve different levels of target sparsity.

In the second candidate architecture, we compress branches using low rank matrix factorization (MF) by decomposing the weight matrices of each RNN layer into two lower dimensional weight matrices via the singular value decomposition (SVD). These methods have been used for model compression in many domains \cite{Andrews1976,Aharon2006,Rufai2014,McGivney2014} including deep learning and speech recognition \cite{Prabhavalkar2016,Kim2019,Swaminathan2020}. In this method, each weight matrix $\boldsymbol{W}^{ \matdimx \times \matdimy}$ in an AmRNN layer is decomposed into two weight matrices $\boldsymbol{P_{1}}=[ \,P_1^1,\ldots,P_1^\matdimx] \,$ and $\boldsymbol{P_2}=[ \,P_2^1,\ldots,P_2^\matdimy] \,$. To achieve varied compute costs across branches, we choose ranks $\textbf{r}=\{r_1,\ldots,r_N\} \leq \min(\matdimx,\matdimy)$ and truncate each weight matrix such that for the $n^{th}$ branch, only the first $r_n$ column vectors in $\boldsymbol{P_{1}}$ and $\boldsymbol{P_{2}}$ are used, e.g., for input $x$ and output $y$,
\noindent
\begin{align}\label{eqn:mat_fac}
	\noindent
	y= [ \,P_1^1,\ldots,P_1^{r_n}] \cdot \left ([ \,P_2^1,\ldots,P_2^{r_n}]^\top x \right )
\end{align}
\noindent
When using $r < (\matdimx \cdot \matdimy)/(\matdimx+\matdimy)$, the number of multiplications is less than when using the original matrix $\boldsymbol{W}$.
Given that the weights for each branch are acquired from the same decomposition, MF lends itself to weight sharing between branches to prevent additional memory overhead.

Both sparse pruning and MF introduce the ability to take advantage of pre-training.
Using a pre-trained RNN-T, one can extract the encoder and apply either compression method to obtain AmRNN branches and reconstruct the model using the remainder of the RNN-T network unchanged.
The model can then be fine-tuned to allow for structural adaptation.
Such a feature allows AmNets to be applied to existing ASR models without having to retrain from scratch.

\textit{
	\begin{table*}[th!]\label{sec:results}
		\centering
		\caption{Baseline RNN-T and amortized RNN-T results. Compression ratio column indicates the overall compression for any given architecture. When two numbers are present in any column, the first indicates the slow (S) branch and the second indicates the fast (F) branch for AmNets. Params column indicates total number of parameters in the encoder network including the arbitrator.}
		\label{tab:results}
		\begin{tabular}{lllrrrrr}
			\toprule
			\textbf{Model} & 
			\textbf{Params} &
			\textbf{Compr. Ratio} & 
			\textbf{WER}	& 
			\textbf{Branch Ratios (S/F)} & 
			\multicolumn{2}{c}{\textbf{FLOPs/Frame}} & 
			\textbf{Latency (ms)} \\ \hline
			Baseline											& 42.7M				& 0\%     				& 8.5	    	& -  						& 42.7M 			& (0.0\%)				& 6154			\\
			Sparse Baseline 									& 36.3M				& 15\%              	& 8.7    		& -  						& 37.1M 			& (-13.1\%)  			& 5231			\\
			Sparse AmNet $\mathcal{L}_{\text{avg}}$				& 45.3M				& {15\%, 80\%}  		& 8.6	 		& 57\% / 43\%				& 27.7M 			& (-35.2\%)				& 681			\\
			Sparse AmNet $\mathcal{L}_{\text{avg}}$	Biased		& 45.3M				& {15\%, 80\%}  		& 9.2	 		& 48\% / 52\%				& 25.3M 			& (-40.8\%)				& 234			\\
			\textbf{Sparse AmNet $\mathcal{L}_{\text{amr}}$}	& \textbf{45.3M}	& \textbf{15\%, 80\%}  	& \textbf{8.7}	& \textbf{48\% / 52\%} 		& \textbf{25.3M}	& \textbf{(-40.8\%)}	& \textbf{1.87}\\
			MF Baseline  										& 32.3M				& 35\% 			  		& 8.5       	& -          				& 30.4M 			& (-28.8\%)				& 3410			\\
			MF AmNet $\mathcal{L}_{\text{avg}}$					& 33.0M				& {35\%, 60\%}  		& 8.6	 		& 38\% / 62\%				& 24.6M 			& (-42.3\%)				& 433			\\
			MF AmNet $\mathcal{L}_{\text{avg}}$	Biased			& 33.0M				& {35\%, 60\%}  		& 9.1	 		& 23\% / 77\%				& 23.2M 			& (-45.6\%)				& 29.4			\\
			\textbf{MF AmNet $\mathcal{L}_{\text{amr}}$}	& \textbf{33.0M}	& \textbf{35\%, 60\%}  	& \textbf{8.6}	& \textbf{23\% / 77\%}		& \textbf{23.2M}	& \textbf{(-45.6\%)}	& \textbf{9.00}\\
			\bottomrule
		\end{tabular}
	\end{table*}
}

\section{Amortized Latency Loss}
\label{sec:am_latency_loss}
\begin{figure}[t!]
	\vspace{-2.0mm}
	\includegraphics[width=\linewidth]{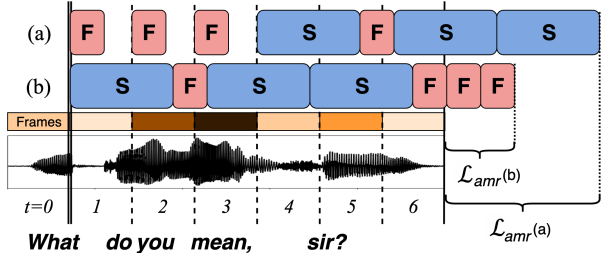}
	\vspace{-2.0mm}
	\caption{\small Amortized latency loss over two cost sequences. Both use the same number of fast (F) and slow (S) decisions and result in the same average cost, yet the (a) sequence often waits for new frames to arrive while (b) yields a lower latency. The amortized latency loss accounts for this difference.}
	\vspace{-4.0mm}
	\label{fig:amortized_loss}
\end{figure}

We now detail two implementations of $\mathcal{L}_{\text{compute}}$, which regularizes the decisions of the arbitrator.
For simplicity, we denote $\computecost_t$ as the compute cost for the frame at timestep $t$ (determined by the arbitrator's decisions) in floating point operations (FLOPs). We consequently define the loss as a function over the sequence costs, $\mathcal{L}_{\text{compute}} \left( \computecost_1, \dots, \computecost_T \right)$.
A straight forward approach for accounting the loss is to consider the total amount of computation or a length normalized mean amount of computation used across the sequence.

\noindent
\begin{align}
	\noindent
	\mathcal{L}_{\text{avg}} = \frac{1}{T} \sum_{i = 1}^T {\computecost_i}
\end{align}
\noindent
This type of regularization is common for other variable compute approaches \cite{Tyagi2020,Bengio2015b,Graves2016}.
However, in interactive settings where response latency is critical, such as voice assistants, the average compute loss serves as more of a proxy regularizer than the true objective of minimizing response latency.
To understand why a decrease in average compute cost of an audio sequence does not necessarily translate into a latency reduction, consider the two cost sequences in Figure \ref{fig:amortized_loss}: (a) mostly fast (F), inexpensive operations followed by a series of slow (S), expensive ones, and (b) the reverse, expensive followed by inexpensive.
Note that the F operations process frames faster than real-time.
While both sequences result in the same average cost, the first sequence has a greater backlog of frames at the end of the utterance caused by underutilized time when the model waited for the next frame. Ultimately, this results in higher latency. 

To address this distinction, we introduce a new measure, termed \emph{amortized latency loss}, which accounts for the processing performance of the hardware executing the computation, the audio feature frame rate, and the decision sequence of the arbitrator.
We denote $\bandwidth$ as the effective FLOP rate (i.e. FLOPs/s) representing the device processor speed and $\framerate$ as the feature frame rate in frames per second.

Defining $\ell_t$ to be the buffered FLOP backlog (i.e. \emph{lag} accumulated at the $t$-th time step), we express the amortized latency loss $\mathcal{L}_{\text{amr}}$ as the response delay in seconds

\noindent
\begin{align}
	\noindent
	\mathcal{L}_{\text{amr}} = \ell_T/\bandwidth
\end{align}
\noindent
with $\ell_t$ expressed recursively as
\noindent
\begin{align}
\ell_t= \left\{
\begin{array}{ll}
      0 & t = 0 \\
      \max\left\{\ell_{t-1} + \computecost_t - \bandwidth/\framerate, 0\right\} & t > 0
\end{array} 
\right. 
\end{align}
\noindent
The loss is convex over $\left( \computecost_1, \dots, \computecost_T \right)$ and is efficiently computable (along with its derivatives) in linear time using dynamic programming.
The amortized latency loss encourages the arbitrator to schedule in a way that achieves high processor utilization while maintaining a short response latency. In this study, $\mathcal{L}_{\text{amr}}$ is regularized and combined with RNN-T sequence loss to achieve reduced latency without sacrificing accuracy.

\section{Experimental Results}
\label{sec:experiments}
We investigate model accuracy using the LibriSpeech corpus \cite{Panayotov2015}. This data set contains 960 hours of training data, and we evaluate using the ``clean'' test data set. All data is pre-processed using a 64-dimensional log-filterbank energy feature extractor. These feature frames are downsampled by three and stacked with a stride size of two to produce 30-millisecond frames. All models are trained to convergence with a warm-hold-decay learning rate schedule and an Adam optimizer \cite{Kingma2015}. AmNets models are first trained with $\mathcal{L}_{\text{avg}}$ loss, which is followed by a short fine-tuning stage using $\mathcal{L}_{\text{amr}}$ loss.

Our baseline model is a lightweight, unidirectional LSTM-based RNN-T with five 1024-unit LSTM layers in the encoder, two 1024-unit LSTM layers in the decoder, and an additive joint network with no trainable parameters. 
We use a vocabulary of 4097 word pieces. 
The baseline model includes 63.5M parameters with 42.7M belonging to the encoder. 
We use standard RNN-T beam search decoding with a width of 16 to obtain word error rate (WER) measurements as our accuracy metric.\footnote{We choose our baseline architecture because it has been successfully tested in on-device, low-latency streaming settings while delivering high accuracy on our internal data. Note that more advanced methods such as language model rescoring \cite{Li2020-a}, bidirectional and frequency LSTMs \cite{Graves2013-BiLSTM, Li2016}, and attention-based transducers \cite{Li2020-b, Zhao2020, Tian2019} exhibit superior WER on the LibriSpeech data set.}

In addition to the overall baseline model, we include baseline models for sparsity and MF that use compression ratios of 15\% and 35\%, respectively. 
These ratios were chosen by continuously increasing the compression for both methods until a development set WER degradation was observed; all baselines achieved 8.5\% WER on LibriSpeech development clean data. 
All AmNet models build upon the baseline model by substituting the encoder LSTM layers with a two-branched AmRNN consisting of five stacked LSTM cells per branch and 1024 units per cell. 
The Sparse AmNets use slow and fast branches with 15\% and 80\% sparsity, respectively, and the MF AmNets use ranks that result in 35\% and 60\% compression, respectively. 
MF AmNets are seeded with weights acquired from the baseline MF model and then trained for a small number of steps until convergence. 
We also attempted to use pre-trained weights for the sparse AmNets but found that training sparse models from scratch yields better results. 
Sparse model branches are pruned with $\Delta \steps=100$ steps for the first 30\% of training using random branch sampling with a 50/50 slow/fast ratio. 
After pruning, Gumbel softmax sampling is activated and used for the remainder of training with $\tau$ linearly annealed from 1.0 to 0.5. 

All arbitrator networks consist of two 128-unit LSTM layers and a single two-unit output projection layer. The arbitrator is pre-trained on a task to classify frames as high or low entropy. Binary targets for the arbitrator pre-training are generated by thresholding a baseline model's output distribution entropy with a threshold chosen between 1.5-2.5 achieving similar results, yielding 40/60-20/80\% initial branch ratios (slow/fast). For all AmNets, no more than 2.5\% of the trainable encoder parameters are attributed to the arbitrator network.

For all models, we report WER, average compute cost, and latency. Compute cost is reported in terms of encoder FLOPs per frame including arbitrator cost. Latency is reported in milliseconds and is the average across all test utterances. To estimate latency, we use a single-threaded compute performance of 650 million FLOPs per second, chosen for two reasons. First, this level of performance is less than half the requirement for a device to maintain real-time (i.e. zero latency) processing when using the baseline model. Second, this level is commensurate with the compute power available when using off-the-shelf embedded systems in the same class as Echo devices.  Note that when designing a model for on-device inference, one would know the compute performance \emph{a priori} and would instead design encoder branch architectures based on device capabilities.

The accuracy, compute cost, and latency for each model is shown in Table \ref{tab:results}. During the average compute cost loss ($\mathcal{L}_{\text{avg}}$) learning phase, we use a regularization factor to strike a balance between RNN-T loss and compute cost loss. This balance is achieved by finding the point at which the AmNets branch ratios use the fast branch to the greatest extent without incurring significant accuracy degradation. For both sparsity and MF, the AmNets trained purely with $\mathcal{L}_{\text{avg}}$ significantly outperform their baseline models with respect to compute cost and latency without accuracy degradation. In this case, MF outperformed sparsity by both measures due to the higher fast branch ratio for MF. Note that all sparse AmNets outperformed the sparse baseline with respect to both FLOPS per frame and latency despite having significantly more parameters.

During the amortized latency loss ($\mathcal{L}_{\text{amr}}$) fine-tuning stage, we found that the branch ratios shifted towards the fast branch--from 43\% to 52\% for sparsity and 62\% to 77\% for MF. To illustrate that the latency improvements realized by tuning with $\mathcal{L}_{\text{amr}}$ are not purely the result of this ratio shift, we increase the $\mathcal{L}_{\text{avg}}$ regularization to achieve \emph{biased} branch ratios that match those observed when using $\mathcal{L}_{\text{amr}}$. While the biased $\mathcal{L}_{\text{avg}}$ models further reduce compute cost and latency, they also incur noticeable accuracy reductions caused by forcing the arbitrator to an unbalanced accuracy-latency trade-off. In contrast, the $\mathcal{L}_{\text{amr}}$ models achieve nearly real-time latency with a $>$99.9\% reduction over all baselines while achieving baseline-level accuracy.

It is important to highlight that the biased MF $\mathcal{L}_{\text{avg}}$ AmNet reduced overall latency by a large margin over the biased sparse AmNet model thus yielding a less significant latency improvement than sparsity when moving to $\mathcal{L}_{\text{amr}}$ loss. This phenomenon emphasizes that, while reducing the average compute cost \emph{may} result in a similarly reduced and nearly real-time latency, this is not necessarily guaranteed. Depending on the specific architecture, compression ratios, and device performance, the unbounded $\mathcal{L}_{\text{avg}}$-trained arbitrator decision sequences only promise an overall average compute cost reduction. The $\mathcal{L}_{\text{amr}}$ loss tuning, however, delivers such a latency guarantee for both candidate architectures by enforcing more deliberate branch decisions and truly amortizing costs based on an awareness of the real-time audio backlog.

\section{Conclusion}
\label{sec:conclusion}
We introduced Amortized Networks accompanied by amortized latency loss and applied them to RNN-T for ASR. By dynamically switching between variable-compute encoder branches on a frame-by-frame basis with amortized latency loss, AmNets are able to yield significant reductions in compute cost and latency. AmNets can be applied to a variety of scenarios and are especially well-suited for sequence-to-sequence modeling.

\clearpage
\newpage

\bibliographystyle{IEEEtran}
\bibliography{./mybib}

\begin{thebibliography}{10}
\providecommand{\url}[1]{#1}
\csname url@samestyle\endcsname
\providecommand{\newblock}{\relax}
\providecommand{\bibinfo}[2]{#2}
\providecommand{\BIBentrySTDinterwordspacing}{\spaceskip=0pt\relax}
\providecommand{\BIBentryALTinterwordstretchfactor}{4}
\providecommand{\BIBentryALTinterwordspacing}{\spaceskip=\fontdimen2\font plus
\BIBentryALTinterwordstretchfactor\fontdimen3\font minus
  \fontdimen4\font\relax}
\providecommand{\BIBforeignlanguage}[2]{{%
\expandafter\ifx\csname l@#1\endcsname\relax
\typeout{** WARNING: IEEEtran.bst: No hyphenation pattern has been}%
\typeout{** loaded for the language `#1'. Using the pattern for}%
\typeout{** the default language instead.}%
\else
\language=\csname l@#1\endcsname
\fi
#2}}
\providecommand{\BIBdecl}{\relax}
\BIBdecl

\bibitem{Graves2013}
A.~Graves, A.~R. Mohamed, and G.~Hinton, ``Speech recognition with deep
  recurrent neural networks,'' \emph{Proc. ICASSP}, 2013.

\bibitem{Rao2018}
K.~Rao, H.~Sak, and R.~Prabhavalkar, ``Exploring architectures, data and units
  for streaming end-to-end speech recognition with {RNN}-transducer,'' in
  \emph{Proc. ASRU Workshop}, Jan. 2018.

\bibitem{Kannan2019}
A.~Kannan, A.~Datta, and T.~N.~S. et~al., ``Large-scale multilingual speech
  recognition with a streaming end-to-end model,'' in \emph{Proc. Interspeech},
  Sep. 2019.

\bibitem{Sainath2020}
T.~N. Sainath, Y.~He, and B.~L. et~al., ``A streaming on-device end-to-end
  model surpassing server-side conventional model quality and latency,'' in
  \emph{Proc. ICASSP}, 2020.

\bibitem{Hochreiter1997}
S.~Hochreiter and J.~Schmidhuber, ``Long short-term memory,'' \emph{Neural
  Computation}, vol.~9, 1997.

\bibitem{Greff2017}
K.~Greff, R.~K. Srivastava, J.~Koutnik, B.~R. Steunebrink, and J.~Schmidhuber,
  ``{LSTM}: A search space odyssey,'' \emph{IEEE Transactions on Neural
  Networks and Learning Systems}, vol.~28, 2017.

\bibitem{Jia2017}
Y.~K. Jia, Z.~Wu, Y.~Xu, D.~Ke, and K.~Su, ``Long short-term memory projection
  recurrent neural network architectures for piano's continuous note
  recognition,'' \emph{Journal of Robotics}, 2017.

\bibitem{Lei2018}
T.~Lei, Y.~Zhang, S.~I. Wang, H.~Dai, and Y.~Artzi, ``Simple recurrent units
  for highly parallelizable recurrence,'' in \emph{Proc. EMNLP}, 2018.

\bibitem{Zhu2018}
M.~H. Zhu and S.~Gupta, ``To prune, or not to prune: Exploring the efficacy of
  pruning for model compression,'' in \emph{6th International Conference on
  Learning Representations, ICLR - Workshop Track Proceedings}, 2018.

\bibitem{Shangguan2019}
Y.~Shangguan, J.~Li, L.~Qiao, R.~Alvarez, and I.~McGraw, ``Optimizing speech
  recognition for the edge,'' \emph{arXiv preprint abs/1909.12408}, 2019.

\bibitem{Pang2018}
R.~Pang, T.~N. Sainath, and R.~P. et~al., ``Compression of end-to-end models,''
  in \emph{Proc., Interspeech}, Sep. 2018.

\bibitem{Alvarez2016}
R.~Alvarez, R.~Prabhavalkar, and A.~Bakhtin, ``On the efficient representation
  and execution of deep acoustic models,'' in \emph{Proc. Interspeech}, Sep.
  2016.

\bibitem{Nguyen2020}
H.~D. Nguyen, A.~Alexandridis, and A.~Mouchtaris, ``Quantization aware training
  with absolute-cosine regularization for automatic speech recognition,'' in
  \emph{Proc. Interspeech}, Oct. 2020.

\bibitem{Panchapagesan2020}
S.~Panchapagesan, D.~S. Park, C.~C. Chiu, Y.~Shangguan, Q.~Liang, and
  A.~Gruenstein, ``Efficient knowledge distillation for {RNN}-transducer
  models,'' \emph{arXiv preprint abs/2011.06110}, 2020.

\bibitem{Tyagi2020}
A.~Tyagi, V.~Sharma, R.~Gupta, L.~Samson, N.~Zhuang, Z.~Wang, and B.~Campbell,
  ``Fast intent classification for spoken language understanding systems,'' in
  \emph{Proc. ICASSP}, May 2020.

\bibitem{Bengio2015b}
E.~Bengio, P.-L. Bacon, J.~Pineau, and D.~Precup, ``Conditional computation in
  neural networks for faster models,'' \emph{arXiv preprint abs/1511.06297},
  Nov. 2015.

\bibitem{Graves2016}
A.~Graves, ``Adaptive computation time for recurrent neural networks,''
  \emph{arXiv preprint abs/1603.08983}, 2016.

\bibitem{Jernite2016}
Y.~Jernite, E.~Grave, A.~Joulin, and T.~Mikolov, ``Variable computation in
  recurrent neural networks,'' in \emph{International Conference on Learning
  Representations, ICLR}, Nov. 2017.

\bibitem{Bolukbasi2017}
T.~Bolukbasi, J.~Wang, O.~Dekel, and V.~Saligrama, ``Adaptive neural networks
  for efficient inference,'' in \emph{34th International Conference on Machine
  Learning, ICML}, vol.~2, 2017.

\bibitem{Jang2017}
E.~Jang, S.~Gu, and B.~Poole, ``Categorical reparameterization with
  gumbel-softmax,'' in \emph{5th International Conference on Learning
  Representations, ICLR - Conference Track Proceedings}, 2017.

\bibitem{Gumbel1960}
E.~J. Gumbel, ``Bivariate exponential distributions,'' \emph{Journal of the
  American Statistical Association}, vol.~55, 1960.

\bibitem{Macoskey21}
J.~Macoskey, G.~Strimel, and A.~Rastrow, ``Bifocal neural {ASR}: Exploiting
  keyword spotting for inference optimization,'' in \emph{Proc. ICASSP}, Jun.
  2021.

\bibitem{He2019}
Y.~He, T.~N. Sainath, and R.~P. et~al., ``Streaming end-to-end speech
  recognition for mobile devices,'' in \emph{Proc. ICASSP}, May 2019.

\bibitem{Wu2020-Sparsity}
Z.~Wu, D.~Zhao, Q.~Liang, J.~Yu, A.~Gulati, and R.~Pang, ``Dynamic sparsity
  neural networks for automatic speech recognition,'' \emph{arXiv preprint
  abs/2005.10627}, 2020.

\bibitem{Hastie2015}
T.~Hastie, R.~Tibshirani, and M.~Wainwright, \emph{Statistical learning with
  sparsity: The lasso and generalizations}.\hskip 1em plus 0.5em minus
  0.4em\relax CRC Press, 2015.

\bibitem{Zhou2019}
X.~Zhou, Z.~Du, and S.~Z. et~al., ``Addressing sparsity in deep neural
  networks,'' \emph{IEEE Transactions on Computer-Aided Design of Integrated
  Circuits and Systems}, vol.~38, 2019.

\bibitem{Gale2019}
T.~Gale, E.~Elsen, and S.~Hooker, ``The state of sparsity in deep neural
  networks,'' \emph{arXiv preprint abs/1902.09574}, 2019.

\bibitem{Hebiri2020}
M.~Hebiri and J.~Lederer, ``Layer sparsity in neural networks,'' \emph{arXiv
  preprint abs/2006.15604}, 2020.

\bibitem{Andrews1976}
H.~C. Andrews and C.~L. Patterson, ``Singular value decomposition ({SVD}) image
  coding,'' \emph{IEEE Transactions on Communications}, vol.~24, 1976.

\bibitem{Aharon2006}
M.~Aharon, M.~Elad, and A.~Bruckstein, ``{K-SVD}: An algorithm for designing
  overcomplete dictionaries for sparse representation,'' \emph{IEEE
  Transactions on Signal Processing}, vol.~54, 2006.

\bibitem{Rufai2014}
A.~M. Rufai, G.~Anbarjafari, and H.~Demirel, ``Lossy image compression using
  singular value decomposition and wavelet difference reduction,''
  \emph{Digital Signal Processing: A Review Journal}, vol.~24, 2014.

\bibitem{McGivney2014}
D.~F. McGivney, E.~Pierre, D.~Ma, Y.~Jiang, H.~Saybasili, V.~Gulani, and M.~A.
  Griswold, ``{SVD} compression for magnetic resonance fingerprinting in the
  time domain,'' \emph{IEEE Transactions on Medical Imaging}, vol.~33, 2014.

\bibitem{Prabhavalkar2016}
R.~Prabhavalkar, O.~Alsharif, A.~Bruguier, and L.~McGraw, ``On the compression
  of recurrent neural networks with an application to {LVCSR} acoustic modeling
  for embedded speech recognition,'' in \emph{Proc. ICASSP}, May 2016.

\bibitem{Kim2019}
H.~Kim, M.~U.~K. Khan, and C.~M. Kyung, ``Efficient neural network
  compression,'' in \emph{Proceedings of the IEEE Computer Society Conference
  on Computer Vision and Pattern Recognition}, Jun. 2019.

\bibitem{Swaminathan2020}
S.~Swaminathan, D.~Garg, R.~Kannan, and F.~Andres, ``Sparse low rank
  factorization for deep neural network compression,'' \emph{Neurocomputing},
  vol. 398, 2020.

\bibitem{Panayotov2015}
V.~Panayotov, G.~Chen, D.~Povey, and S.~Khudanpur, ``Librispeech: An {ASR}
  corpus based on public domain audio books,'' in \emph{Proc. ICASSP}, 2015.

\bibitem{Kingma2015}
D.~P. Kingma and J.~L. Ba, ``Adam: A method for stochastic optimization,'' in
  \emph{3rd International Conference on Learning Representations, ICLR -
  Conference Track Proceedings}, 2015.

\bibitem{Li2020-a}
J.~Li, R.~Zhao, and Z.~M. et~al., ``Developing {RNN-T} models surpassing
  high-performance hybrid models with customization capability,'' in
  \emph{Proc. Interspeech}, Oct. 2020.

\bibitem{Graves2013-BiLSTM}
A.~Graves, N.~Jaitly, and A.~R. Mohamed, ``Hybrid speech recognition with deep
  bidirectional {LSTM},'' in \emph{Proc. ASRU}, 2013.

\bibitem{Li2016}
J.~Li, A.~Mohamed, G.~Zweig, and Y.~Gong, ``Exploring multidimensional {LSTM}s
  for large vocabulary {ASR},'' in \emph{Proc. ICASSP}, May 2016.

\bibitem{Li2020-b}
J.~Li, Y.~Wu, Y.~Gaur, C.~Wang, R.~Zhao, and S.~Liu, ``On the comparison of
  popular end-to-end models for large scale speech recognition,'' in
  \emph{Proc. Interspeech}, Oct. 2020.

\bibitem{Zhao2020}
Y.~Zhao, C.~Ni, C.~C. Leung, S.~Joty, E.~S. Chng, and B.~Ma, ``Cross attention
  with monotonic alignment for speech transformer,'' in \emph{Proc.
  Interspeech}, Oct. 2020.

\bibitem{Tian2019}
Z.~Tian, J.~Yi, J.~Tao, Y.~Bai, and Z.~Wen, ``Self-attention transducers for
  end-to-end speech recognition,'' in \emph{Proc. Interspeech}, Sep. 2019.

\end{thebibliography}

\end{document}